\documentclass{article}
\usepackage{hiph-preprint}
\usepackage{graphicx}
\usepackage{amssymb}
\volnumber{22} \issuenumber{1} \edyear{2005}                             %|
\frompage{000} \topage{000}                                              %|
\recrevdate{1 January 2005}                                              %|
\def\rightmark{Mass generation in coalescence}
\def\leftmark{T. Peitzmann}
 
\title{Mass generation in coalescence - effects on hadron spectra} 
\authors{ 
{Thomas Peitzmann$^1$%
\index{Peitzmann, T.} % Abbreviated names of the author(s),
}\\[2.812mm]
{\normalsize
\hspace*{-8pt}$^1$ Utrecht University/NIKHEF, 
3508TA Utrecht, The Netherlands\\[0.2ex] 
}}
 
\abstract{Different scenarios for the creation of constituent mass in the hadron formation process are discussed. Effects of these may be observable in hadron momentum spectra.}

 
\makeindex
\begin{document}
 
\maketitle

\section{Introduction}\label{intro}

It has been proposed that the particle species dependences observed in the intermediate $p_T$ regime ($2 - 6 \, \mathrm{GeV}/c$) of heavy ion collisions \cite{phenix:quench,star:quench,star:flow,star:flowlam,phenix:bar} may be explained by a collective production mechanism, namely \textit{recombination} or \textit{coalescence}  \cite{fries:reco,greco:reco}.
In most coalescence models hadrons are assumed to form from essentially collinear partons. The overlap function is sometimes simply assumed to be a delta function, in some cases finite widths have been used, usually very small in the transverse direction and assuming an $x$ distribution like one expects to see in the final state hadron, such that the partons do not have to undergo a change in momentum when forming a hadron. There is no way that this kind of coalescence could produce hadrons with the same quark content but different masses, if one doesn't assume that quarks of the same flavor may carry different constituent masses.
Hadrons of different mass may be produced, if one allows for relative momentum of the partons at the moment of coalescence. Then a natural step is to allow quarks to have smaller masses - may be even their current masses - allowing them to also form pions. It may be useful to relax the assumption of quarks having their constituent mass and investigate what difference a change in the quark masses makes.
 
%One may envisage two different extreme scenarios for the generation of masses:
%\begin{enumerate}
%\item Quarks may obtain their constituent masses early before confinement sets in. These \textit{dressed} quarks would then form hadrons by coalescence.
%\item Quarks remain massless until confinement. Constituent masses are generated at the moment of hadron creation, i.e. coalescence.
%\end{enumerate}
%Although the real situation may be somewhat intermediate between the two extremes given above, I will for simplicity concentrate on discussing these extreme cases. In both cases gluons are ignored. 
One may envisage two different extreme scenarios for the generation of masses:\\
1.~Quarks may obtain their constituent masses early before confinement sets in. These \textit{dressed} quarks would then form hadrons by coalescence.\\
2.~Quarks remain massless until confinement. Constituent masses are generated at the moment of hadron creation, i.e. coalescence.\\
Although the real situation may be somewhat intermediate between the two extremes given above, I will for simplicity concentrate on discussing these extreme cases. In both cases gluons are ignored.

In the first scenario hadrons are made by joining collinear constituents with the same momentum fraction, and one can estimate the spectrum by boosting the thermal spectra of the constituents simultaneously, which is essentially the same as calculating the boosted spectrum of the final hadron \cite{Schn93}:
\begin{equation}
\frac{1}{2 \pi p_T} \frac{dN}{d p_T} = 
 C \cdot m_T \cdot I_0\left( \frac{p_T \sinh \eta_T}{T} \right) 
K_1\left( \frac{m_T \cosh \eta_T}{T} \right).
\label{eq1}
\end{equation} 

In the second scenario, the thermal movement has to be used at least partly to create the mass at the moment of coalescence. Approximately one may treat this as the simultaneous boost of massless partons, whose transverse momenta will add up to the hadron transverse momentum, but the mass generation produces another Boltzmann penalty factor:
\begin{equation}
\frac{1}{2 \pi p_T} \frac{dN}{d p_T} = C \cdot e^{-m/T} \cdot p_T \cdot I_0\left( \frac{p_T \sinh \eta_T}{T} \right) 
K_1\left( \frac{p_T \cosh \eta_T}{T} \right).
\label{eq2}
\end{equation} 
In addition to the Boltzmann factor the only difference is a replacement of all occurrences of $m_T$ on the right hand side by $p_T$. The normalization factor, which contains the spatial density, spin multiplicities and a transition probability for forming the hadron, must not necessarily be identical for the different cases, but I will assume this for simplicity here.

\section{Results}\label{results}

\begin{figure}[htb]
\resizebox{0.5\textwidth}{!}{\includegraphics{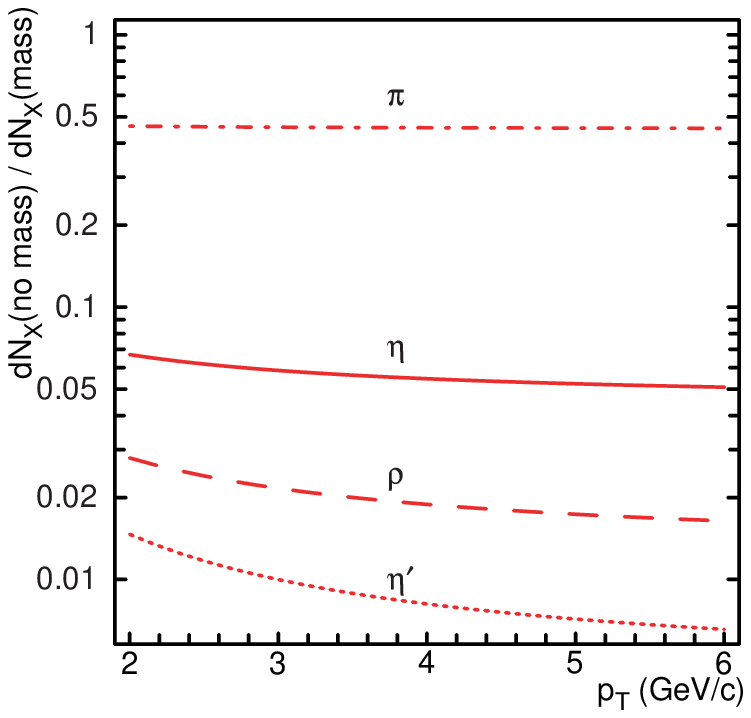}}
\resizebox{0.5\textwidth}{!}{\includegraphics{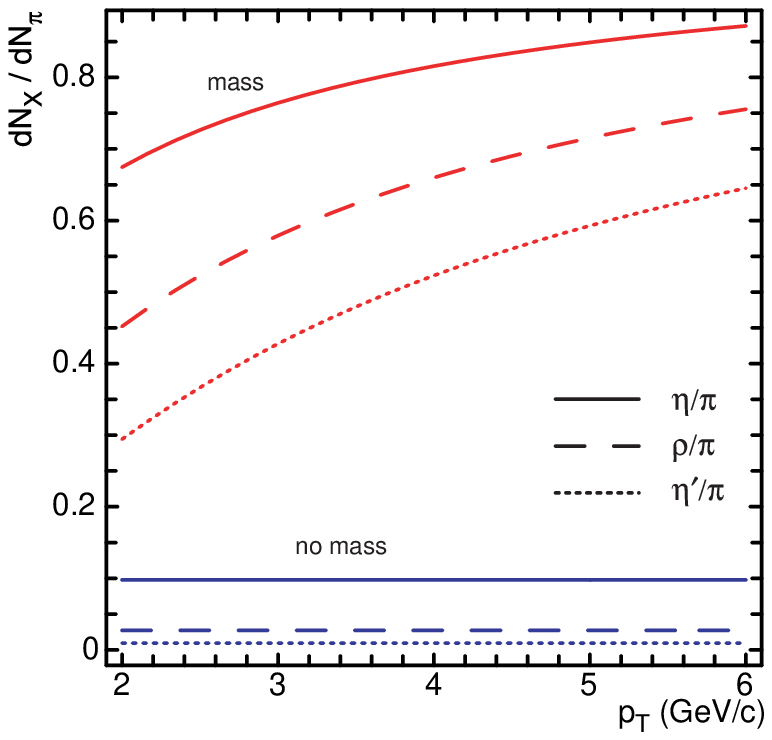}}
\caption{Comparison of production from massless partons (\textit{no mass}) to the production from massive partons (\textit{mass}) for $T = 175 \, \mathrm{MeV}$ and $\eta_{T} = 0.3$. The different spin multiplicities for the vector meson are not accounted for.
Left: Ratio of transverse momentum spectra of different mesons for the two scenarios. Right: Ratio of transverse momentum spectra of different mesons to those of pions. The two groups of curves show results for the different scenarios.
\protect\label{fig1}}
\end{figure}
The left side of Fig.~\ref{fig1} shows the ratio of the momentum distributions using equation~\ref{eq2} (no mass) to those for equation~\ref{eq1} (mass) for mesons of different masses. I have chosen a temperature $T = 175 \, \mathrm{MeV}$ and a flow velocity $\eta_{T} = 0.3$. The major effect is a \textit{hadrochemical} suppression, if the masses are created at the transition. The suppression is not as strong at the lower $p_T$ end for heavier mesons, because the conversion from $m_T$ to $p_T$ causes a reduction in yield for heavier particles for the case of equation~\ref{eq1}. Certainly, mass production at earlier stages to produce constituent quarks would also "consume" energy, but the system would have time to equilibrate the massive quarks up to the phase transition temperature.

As can be see already from the left side of Fig.~\ref{fig1} the two different scenarios investigated here will lead to significantly different behavior of particle ratios in this transverse momentum range. More explicitly this can be seen on the right side of Fig.~\ref{fig1}, where the ratios of heavier mesons to pions are shown. In the \textit{mass}-scenario heavier mesons are suppressed at low $p_T$ while the ratio increases for higher $p_T$ similar to ordinary thermal hadron spectra. In the \textit{no mass}-scenario the ratio is just reduced by the Boltzmann factors containing the masses. There are no other dependences on the hadron masses in equation~\ref{eq2}, and so the ratios are independent of $p_T$.
I will not discuss in detail the dependence on the other parameters -- it turns out that the influence of the flow velocity is small for intermediate $p_T$, and, while there is a dependence on the temperature, a particular value is suggested by predictions for the phase transition temperature.
 
\begin{figure}[htb]
\resizebox{0.55\textwidth}{!}{\includegraphics{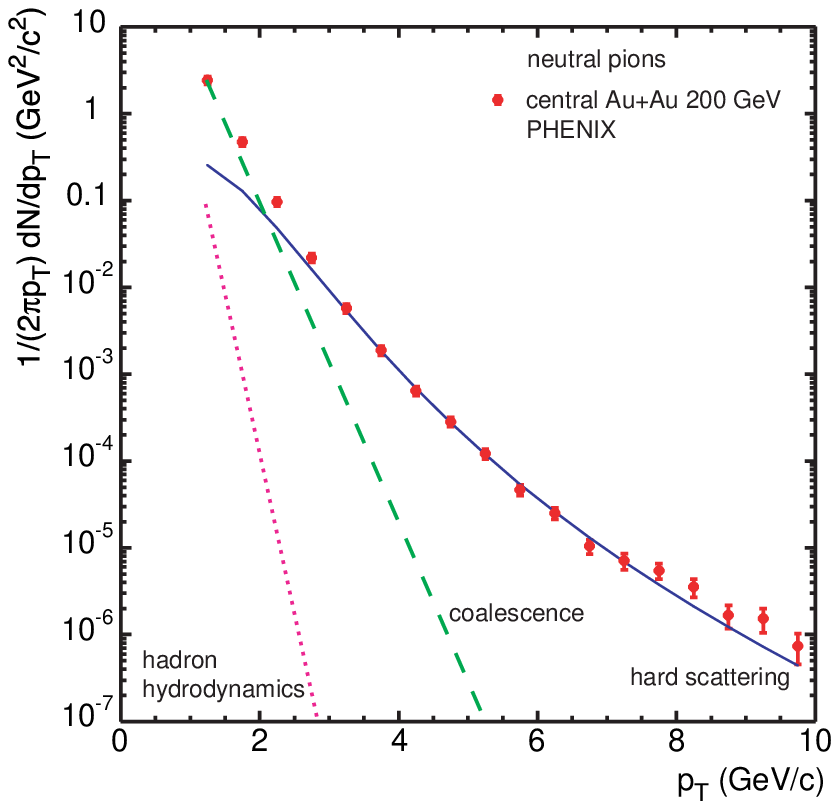}}
\resizebox{0.45\textwidth}{!}{\includegraphics{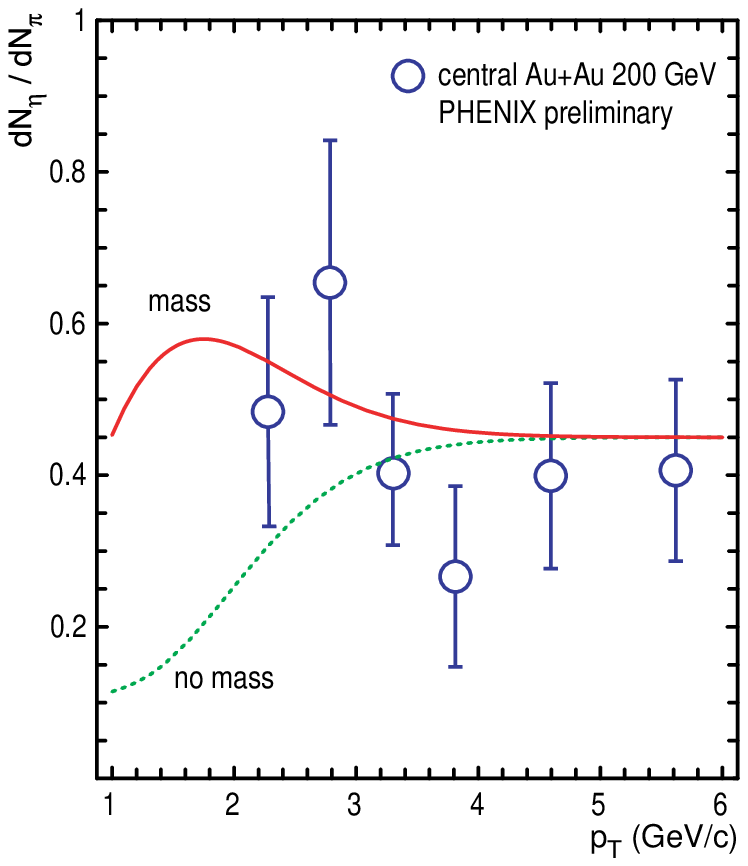}}
\caption{Left: The different components of the toy model calculation compared to the $\pi^0$ spectrum from central Au+Au collisions as measured by PHENIX \cite{phenix:pi0}.
Right: The $\eta/\pi^0$ ratio as predicted from the two different coalescence scenarios compared to preliminary data measured by PHENIX \cite{phenix:etapi}.
\protect\label{fig2}}
\end{figure}

The observable effects may be obscured by the admixtures of hadrons from other production mechanisms. To investigate this, the parameterizations from equations \ref{eq1} and \ref{eq2} have been fitted tigether with a power law for high $p_T$ to the neutral pion spectra in central Au+Au collisions (see left side of Fig.~\ref{fig2}).The $\eta$ spectrum can then be obtained using a ratio $\eta/\pi^0 = 0.45$ for the power law and the expected mass scaling for the coalescence estimates.The resulting $\eta/\pi^0$ ratio for the two scenarios is displayed on the right side of Fig.~\ref{fig2}. Below $p_T = 3 \, \mathrm{GeV}/c$ the two scenarios yield considerably different estimates. Also shown are preliminary data from the PHENIX experiment \cite{phenix:etapi}. Within the present experimental uncertainties and $p_T$ reach the data do not allow to distinguish between the two scenarios. 

\section{Conclusions}\label{concl}

In a simplified model the consequences of different scenarios for the generation of hadron masses in coalescence have been discussed. If the hadron mass is generated only at the phase transition, one should observe a suppression of heavier mesons with similar quantum numbers, compared to a scenario where quarks carry the constituent mass at the moment of coalescence. This should lead to observable consequences even when other production mechanism are taken into account. Present data do not allow to distinguish between the scenarios.

The model discussed here is, however, very limited. In view of the interesting information on the process of mass generation, an implementation of these different scenarios in a realistic coalescence calculation would be desirable.

\vfill\eject

\begin{thebibliography}{99}

\bibitem{phenix:quench}
K.~Adcox et al.  [PHENIX Coll.],
 {\it Phys.\ Rev.\ Lett.\ } {\bf 88}, 022301 (2002).

\bibitem{star:quench}
 C.~Adler et al.  [STAR Coll.],
 {\it Phys.\ Rev.\ Lett.\ }  {\bf 90}, 082302 (2003).

\bibitem{star:flow}
 C.~Adler et al.  [STAR Coll.],
{\it Phys.\ Rev.\ Lett.\ }  {\bf 90}, 032301 (2003).

\bibitem{star:flowlam}
 C.~Adler et al.  [STAR Coll.],
 {\it Phys.\ Rev.\ Lett.\ }  {\bf 92}, 052302 (2004).

\bibitem{phenix:bar}
S.S.~Adler et al.  [PHENIX Coll.],
{\it Phys.\ Rev.\ Lett.\ } {\bf 91}, 172301 (2003).

\bibitem{fries:reco}
R.~J.~Fries et al.,
{\it Phys.\ Rev.\ Lett.\ } {\bf 90}, 202303 (2003).

\bibitem{greco:reco}
V.~Greco, C.~M.~Ko and P.~Levai,
{\it Phys.\ Rev.\ Lett.\ } {\bf 90}, 202302 (2003).

\bibitem{Schn93}
E.~Schnedermann, J.~Sollfrank, and U.~Heinz, Phys. Rev. C {\bf 48}, 2462 (1993).

\bibitem{phenix:pi0}
S.S.~Adler et al.  [PHENIX Coll.],
{\it Phys.\ Rev.\ Lett.\ } {\bf 91}, 072301 (2003).

\bibitem{phenix:etapi}
 H.~Busching [PHENIX Collaboration], Eur. Phys. J. C {\bf 43}, 303 (2005).

\end{thebibliography}
\end{document}